# An UWB Hemispherical Vivaldi Array

Carl Pfeiffer and Jeffrey Massman

*Abstract*— **We report the first conformal ultra-wide band (UWB) array on a doubly curved surface for wide angle electronic scanning. We use a quadrilateral mesh as the basis for systematically arraying UWB radiators on arbitrary surfaces. A prototype consisting of a 52 element, dual-polarized Vivaldi array arranged over a 181 mm diameter hemisphere is developed. The antennas and SMP connectors are 3D printed out of titanium to allow for simple fabrication and assembly. We derive the theoretical gain of a hemispherical array based on the antenna size and number of elements. The measured realized gain of the prototype array is within 2 dB of the theoretical value from 2-18 GHz and scan angles out to 120° from the *z*-axis. This field of view is twice that of a planar array with the same diameter in agreement with theory. This work provides a baseline performance for larger conformal arrays that have more uniform meshes. Furthermore, the basic concept can be extended to other UWB radiating elements.**

## I. INTRODUCTION

Significant research and development has been invested in developing high performance, dual-polarized planar arrays that realize ultra-wide bandwidths (UWB) [1], low cross-polarization [2, 3], wide-angle scanning [4], low profile [5, 6] and optimal element spacing [7, 8]. These arrays employ tightly coupled elements arranged in a uniform lattice to realize a small active reflection coefficient over a wide operational bandwidth. The Vivaldi array is a notable example of an UWB planar array that has been extensively utilized due to its simple operation and ability to cover greater than one decade of bandwidth [9, 10]. Planar arrays are attractive because they maximize the antenna gain for given number of elements. However, they suffer from a limited field of view since their projected area falls off as $\cos(\theta)$, where $\theta$ is the angle from broadside. The field of view can be extended using a gimbal. However, gimbals are unattractive because these mechanical systems are slow, bulky, and can wear out over time. Ideally, it would be possible to build an antenna array that covers the surface of a curved platform with UWB radiating elements to enable maximum available gain and field of view at all frequencies of interest.

A myriad of arrays on singly curved surfaces such as a cylinder or cone have been developed to enable wide fields of view [11, 12, 13]. A notable example is [14] where three separate narrowband cylindrical/conical arrays are combined to provide a directivity greater than 17 dB over the entire $4\pi$ steradian field of view. It is conceptually straightforward to wrap an UWB planar array around a singly curved surface such as a cylinder, which makes arrays on singly curved surfaces much easier to design and build than arrays on doubly curved surfaces. For example, a cylindrical array is periodic such that an infinite array that accounts for mutual coupling between neighboring elements can be exactly simulated with periodic boundary conditions. Therefore, the array performance can be optimized through computationally inexpensive unit cell simulations. In contrast, it is generally not possible to periodically tile a doubly curved surface. Hence, it is unclear how to rigorously simulate the unit cell of an array on a doubly curved surface so that mutual coupling between adjacent elements is accurately modelled. The aperiodicity makes UWB array design particularly problematic because UWB arrays typically engineer mutual coupling between antennas to achieve a good active impedance match.

The vast majority of conformal arrays employ narrowband elements with less than one octave of bandwidth. Narrowband radiators can be designed to have low mutual coupling such that the aperture shape has minimal impact on element performance. Most conformal arrays also have relatively large inter-element spacing between antennas (>$0.75\lambda$) to fit the antennas next to each other on the non-periodic lattices that are inherent to doubly curved surfaces. The large inter-element spacing results in low aperture efficiency since grating lobes/sidelobes carry substantial power. For example, [15] and [16] report hemispherical arrays composed of 64 circularly polarized helix or waveguide antennas. The arrays are designed to operate from 8-8.4 GHz and with roughly $0.75\lambda$ element spacing. The arrays are cleverly fed with 16 T/R modules and 4:1 power splitters for efficient utilization of resources. The aperture efficiency is roughly 30%, but this efficiency could likely increase somewhat if more T/R modules are employed. An extreme example of large inter-element spacing is the UWB array of quad-ridge horn antennas pointing spherically outwards reported in [17].

Spherical arrays of patch antennas have also been demonstrated [18, 19]. In [20], relatively wideband microstrip patches with 25% bandwidth are distributed along the surface of a sphere. The minimum spacing between elements is $1.5\lambda$ so grating lobes and low aperture efficiencies are expected. A spherical patch antenna array with reduced height is proposed in [21]. However, the aperture efficiency is still only 25% due to the large inter-element spacing. An alternative approach to

The authors are with US Air Force Research Laboratory, Wright-Patterson Air Force Base, OH 45433 (e-mail: carlpfei@umich.edu, jeffrey.massman.5@us.af.mil).

realizing a wide field of view is to fabricate planar subarrays and integrate them into a three-dimensional frame [22, 23]. However, the seams between the planar subarrays limit the performance.

A common challenge for developing conformal antenna arrays is fabrication. It is customary for every element to be individually fabricated and then combined, which requires a fair amount of undesirable touch labor. Some automated techniques for fabricating conformal antennas by selectively patterning metal on curved surfaces have been developed [24, 25, 26]. However, these fabrication capabilities are best suited for building narrowband antenna arrays. A particularly promising process for fabricating conformal antenna arrays is 3D printing because this process can build complicated UWB antenna geometries quickly and cheaply [10, 27, 28, 29, 30].

Here, we introduce the first UWB antenna array on a doubly curved surface that is capable of electronic scanning. To illustrate the concept, we built a prototype with 52 dual-polarized Vivaldi elements that cover the surface of a hemisphere. The basic concept is to modify planar UWB antenna elements so that they conform to an arbitrary surface. We use commercial metal 3D printing technology to build the array because it allows straightforward fabrication of complex geometries that would be practically impossible or cost prohibitive using conventional subtractive manufacturing techniques such as CNC machining or wire electron discharge machining [31, 32]. Furthermore, we 3D print SMP connectors along with the antenna elements to significantly simplify assembly.

In Section II we discuss the theoretical gain and field of view of a hemispherical array. Then we show how a standard quadrilateral mesh can cover a doubly curved surface with tightly coupled UWB radiating elements, for the first time. In Section III we introduce the antenna element design as well as simulations of the hemispherical array. Section IV discusses the array metal 3D printing fabrication, as well as array testing using a spherical far-field measurement system. Radiation patterns generated through post-processed digital beamforming are in good agreement with simulation. In general, this work is intended to evaluate the feasibility of the proposed concept. We invested minimal effort in design optimization, and therefore, we expect many aspects of the array performance could be improved in the future such as impedance match, cross-polarization, and sidelobe levels.

## II. Hemispherical Array Theory

To demonstrate the UWB conformal array concept, we developed a hemispherical array, which has some attractive wide angle scanning properties. To illustrate this, let us compare the field of view of a hemispherical array with radius $r$ to a planar array on a circular disk with the same radius. Both the hemisphere and disk are oriented such that the z-axis is the symmetrical axis of revolution. We assume the array is large enough such that the gain is proportional to the projected area. It is well known that the projected area of the planar array pointing in the direction $\theta_0$ from normal is given by $\pi r^2 \cos(\theta_0)$. It is easy to show that the projected area of a hemispherical array is given by $\pi r^2 \left(\frac{1}{2} + \frac{\cos(\theta_0)}{2}\right)$, where $\theta_0$ is the angle between the scan direction and the z-axis. The field of view ($FOV$) is the solid angle at which the projected area is above some threshold, and is given by,

$$FOV = 2\pi(1 - \cos(\theta_{max})) \quad (1)$$

for azimuthally symmetric antennas such as the planar disc and hemisphere. Here, $\theta_{max}$ is the maximum scan angle at which the projected area is equal to some threshold (e.g. 3 dB below the peak). Setting the projected areas to be equal for the planar and hemispherical cases, it is straightforward to show that

$$FOV_{hemisphere} = 2\ FOV_{planar} \quad (2)$$

In other words, if we require the gain to be above an arbitrary threshold, the field of view of the hemispherical array will always be twice as large as the field of view of the planar array with the same radius. However, the surface area of the hemispherical array is also twice as large. Therefore, for a given number of radiating elements, a planar array will offer twice the gain but half the field of view as a hemisphere.

The peak gain of a hemispherical array is a function of the radius and number of antenna elements. A hemispherical array with 100% aperture efficiency has gain equal to $4\pi^2 r^2/\lambda^2$, where $\lambda$ is the operating wavelength. The maximum array gain occurs when the unit cell area is $\lambda^2/4$ for square lattice arrays. Reducing the wavelength further creates grating lobes such that the gain remains constant. The minimum wavelength for grating lobe free operation is therefore $\lambda_{min} = r\sqrt{8\pi/N}$, where $N$ is the number of dual-polarized elements covering a hemisphere with surface area $2\pi r^2$. Thus, a hemispherical array with 100% aperture efficiency operating at $\lambda_{min}$ will have a maximum gain ($G_{hemisphere}^{max}$) equal to

$$G_{hemisphere}^{max} = N\pi/2 = G_{planar}^{max}/2 \quad (3)$$

where ($G_{planar}^{max}$) is the gain of a planar array with $N$ elements.

Next, we consider how best to distribute the antenna elements on a hemispherical surface since there are no periodic methods for covering a doubly curved surface. The conceptually simplest approach is to evenly distribute the elements in elevation ($\theta$) and azimuth ($\phi$) in the spherical coordinate system [33]. This approach can result in relatively uniform arrays near $\theta = 90°$. But as $\theta$ approaches the poles at 0° and 180°, the spacing between elements approaches 0, which is not practical. An alternative method is to evenly distribute the antennas along elevation [21]. Then, a unique azimuth spacing is chosen for each elevation angle to help make element spacing more uniform. Other techniques based on polyhedrons, Leopardi's algorithm, and spiral distributions are reported in [34]. An interesting mathematical problem known as the Thomson problem asks how to distribute electrons over the surface of a sphere while minimizing electrostatic potential energy. Placing antennas at the electron locations that solve Thomson's problem represents an optimally uniform lattice that is mostly triangular. These previously published approaches can generally work well for narrowband antennas with high isolation between elements because the elements operate independently of one another. However, UWB arrays require a relatively uniform lattice with high coupling between adjacent elements.

Here, we use commercial meshing tools to cover an arbitrary surface with a quadrilateral mesh. Linearly polarized Vivaldi radiators are then placed along the mesh edges. A quadrilateral mesh is ideal for distributing dual-polarized radiating elements since each mesh face is approximately square. Commercial quadrilateral meshing tools are already optimized to maximize the mesh uniformity and 'squareness', which ensures the antennas can realize a similar active impedance match and orthogonal-port isolation as the planar case. Furthermore, this approach is attractive since it is not limited to spherical geometries. Rather, a quadrilateral mesh can cover an arbitrary surface with UWB radiating elements.

The quadrilateral mesh we use for our hemispherical prototype array is shown in Fig. 1. The mesh has 104 edges, which corresponds to 52 dual-polarized antenna elements. Of course, a mesh with a larger or smaller number of elements could have easily been generated. Our choice of 52 dual-polarized antenna elements represents a compromise between prototype size and performance. This array size is small enough such that cost for fabrication and measurement is relatively low, while the array size is large enough such that finite array edge effects are not too significant. Furthermore, the spacing between antennas affects the operating frequency since grating lobes start to appear when the wavelength is less than twice the antenna spacing. The mesh is generally quite uniform such that every radiating element should behave similarly. Only four vertices are slightly irregular since they are connected to three edges rather than four.

## III. ANTENNA ELEMENT DESIGN

We use a Vivaldi radiator due to its robust operation. Vivaldi antennas are travelling wave structures that employ a balun and a gradual impedance taper from 50 Ω to the free space wave impedance (377 Ω). These arrays can easily generate multiple octaves of bandwidth with very little optimization. Thus, they are quite robust to geometrical variations, which is useful for conformal applications because every element will be slightly different in general. The antennas are constructed from titanium due to its 3D printing accuracy and decent conductivity of $\sigma = 1.82 \times 10^6$ S/m. This conductivity is roughly 30 × lower than copper.

Three different views of the basic antenna element are shown in Fig. 2 (a)-(c). The exact dimensions of the element change depending upon its location in the array. The antenna operates similar to a conventional Vivaldi array where a balun feeds a tapered slot. One modification is the width of the element in Fig. 2(a) increases from roughly 14 mm at the bottom to 34.6 mm at the top, which is necessary to maintain electrical connectivity to neighboring elements along the entire length for this hemispherical lattice. We also modified some features to be more amendable to the metal 3D printing process as described in [10]. One fabrication design rule is parts should grow upwards and outwards at an angle of less than 50° from normal to minimize the number of support structures that need to be manually removed. Our antennas are printed upside down such that they grow in the −z direction in Fig. 2(a) to ensure the part is self-supporting. The edges of the antenna element contain a conical vertex which has a larger diameter near the top of the element to help satisfy the 50° design rule. This is necessary because, there is not a uniform 90° angle between edges attached to each vertex of the quadrilateral lattice in Fig. 1. The antenna is fed with a SMP connector that is also 3D printed to simplify assembly. The connector feeds the symmetric Vivaldi radiating arms using a self-supporting tapered transmission line balun in contrast to a traditional Marchand balun. The Vivaldi arms are gridded to reduce weight and cost. Since the entire array cannot be printed as a single piece, it is important that the antenna element is modular. In other words, the array design is sliced up into separate modules that can be printed independently and then screwed together. The Vivaldi arms are connected to the bottom ground plane with modular shorting posts to ensure that each module comes out of the printer as a single part, as shown in Fig. 2(b).

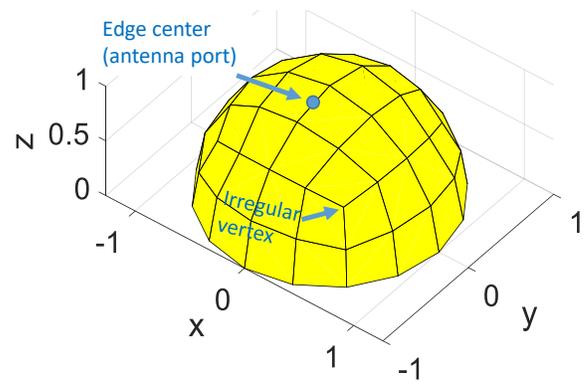

Fig. 1. Quadrilateral mesh that serves as the basis for the antenna element placement on a hemispherical array

As previously mentioned, the exact dimensions of an antenna element depend upon its location in the array because the geometry is not periodic. Fig. 2(d) helps illustrate how the different antenna elements are tweaked within a mesh quadrant to conform to the doubly curved surface. First, the grey cones are added at the locations of the quadrilateral vertices. Next, a Vivaldi element is placed at each edge of the quadrilateral mesh. The dimensions of the Vivaldi antennas near the SMP feed point are all identical. The outer edges of the Vivaldi antenna are expanded/contracted in order to fill the space between the two vertices. In general, the dimensions of a Vivaldi antenna's outer edges minimally affect the performance because the wave is loosely bound to the surface at this point. The overlap between the Vivaldi antennas and the conical vertices helps ensure there is a smooth connection between adjacent antennas, which also improves the 3D printing accuracy. Finally, the conical vertices are hollowed out to reduce weight. Fig. 2(d) corresponds to one of the most distorted quadrants in the mesh because it contains an irregular vertex that is only connected to 3 Vivaldi antennas.

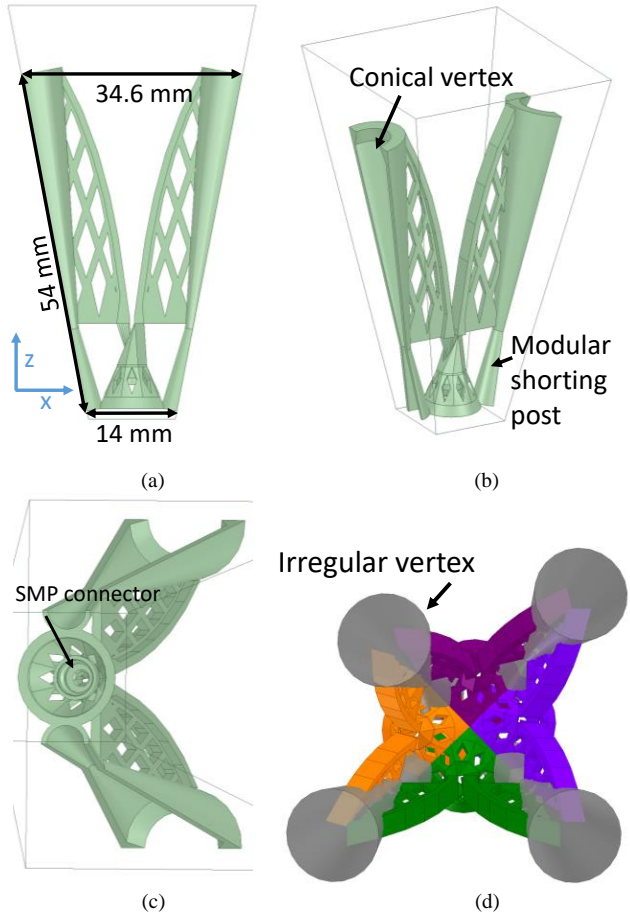

Fig. 2. (a)-(c) Different views of the hemispherical single-pol Vivaldi element. (d) Example illustrating how Vivaldi elements are modified to conform to the doubly curved surface.

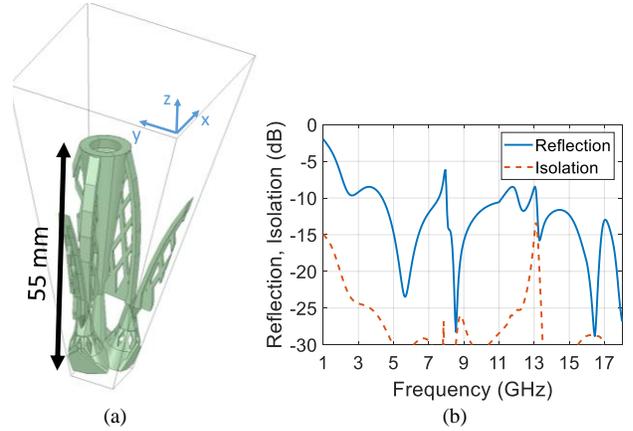

Fig. 3. (a) Dual-polarized Vivaldi unit cell. (b) Simulated active reflection coefficient and orthogonal port isolation.

The simulated 'unit cell' of the dual-polarized antenna element and its performance are shown in Fig. 3(a). The 'unit cell' is simulated in a quasi-infinite array environment. The 4 sides of the unit cell are angled such that they approximate the radius of curvature of the doubly curved antenna geometry. The edges of the simulation domain have periodic boundary conditions with 0° phase delay between opposite sides. This approximates the case where every element is excited in phase. Although this doesn't correspond to the excitation that will be used in the actual array, it does provide a qualitative estimate for the array performance that accounts for mutual coupling.

The active reflection coefficient and orthogonal port isolation are shown in Fig. 3(b). The active reflection for the $x$ and $y$ polarized ports are identical due to the unit cell symmetry. Orthogonal port isolation is defined as the transmission coefficient between the $x$ and $y$ directed Vivaldi antenna ports. In the limit the radius of curvature approaches infinity, the unit cell simulates an infinite planar array pointing towards broadside. The antenna has a decent active impedance match above 2 GHz with reflection below -8 dB for most frequencies. The orthogonal port isolation is also quite low (< -20 dB for most frequencies). There are narrow resonances near 8 GHz and 13 GHz, which are likely due to surface waves. However, the impact of these surface waves is often reduced when the array is finite and not periodic. It should be emphasized that the unit cell in Fig. 3(a) is not optimized for a low reflection coefficient since the simulation only provides a qualitative performance estimate of the hemispherical array. Instead, we simply rely on the fact that Vivaldi radiators generally have a good impedance match when the antenna height is greater than $\lambda/2$.

The simulated radiation efficiency is greater than 95% across the band (1-21 GHz) even though the metal conductivity is 30× lower than that of copper. The Vivaldi antenna has a high radiation efficiency because it is not resonant, has low peak current density, and a moderate electrical length of $3.8\lambda_H$ at the maximum operating frequency.

Our prototype array has 104 ports which corresponds to 52 dual polarized antenna elements. Therefore, (3) suggests the maximum gain equals 19.1 dB. The array is 181.5 mm in diameter, which corresponds to a minimum wavelength of $\lambda_{min} = 126$ mm (4.75 GHz) for grating lobe free operation. The hemispherical array design is shown in Fig. 4(a). We wrote a script that automatically distributes antenna elements along the edges of the quadrilateral mesh in Fig. 1. Then, the array is manually 'sliced up' to create 20 separate modules that can be printed individually. An example module is shown in Fig. 4(b). The seams between modules slice through the antenna vertices and away from the feed. Thus, the seams are located where the current density is generally lowest, which helps minimize the impact on an imperfect electrical contact between adjacent modules. This is important because the modules are simply butted up against one another on the fabricated array which generates a relatively poor electrical connection between them.

## IV. SIMULATION

Planar arrays are generally periodic which implies every element has the same radiation pattern. This fact allows analytic prediction of the array's radiation patterns at different pointing angles, and therefore patterns are often not reported. However, a hemispherical array is not truly periodic. Therefore, it is important to analyze the patterns at various pointing angles on the finite array to evaluate the antenna performance. It is not practical for us to simulate the full hemispherical geometry shown in Fig. 4(a) using a full-wave solver. Therefore, we generated the simplified antenna model shown in Fig. 5 to estimate the performance of this hemispherical design. This simplified model removes most of the subwavelength features of the actual design to reduce the simulation mesh. For example, the tapered transmission line balun is replaced with an ideal lumped port that feeds the symmetric Vivaldi arms. Nevertheless, this model it is still useful for estimating many aspects of the array performance.

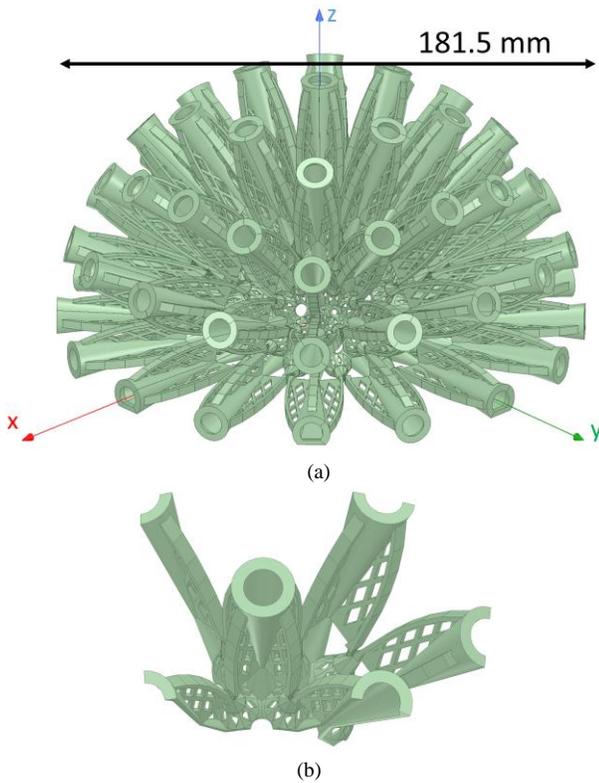

Fig. 4. (a) Designed Hemispherical Vivaldi array. (b) One of the 20 modules comprising the array.

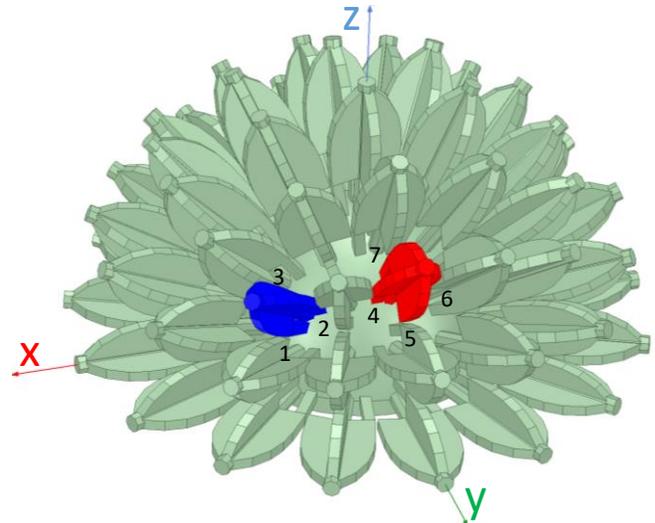

Fig. 5. Simplified array model that is simulated to estimate antenna performance. The scattering parameters of the ports connected to the blue and red vertices are plotted in Fig. 6.

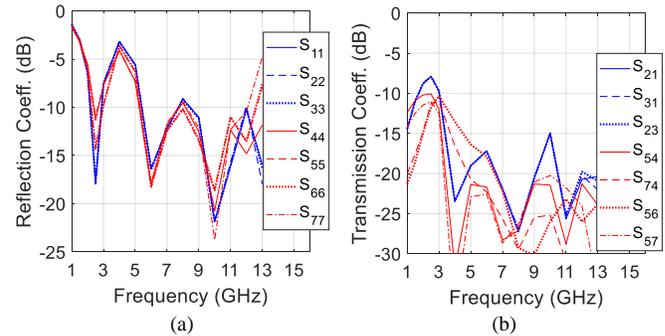

Fig. 6. Reflection coefficient (a) and transmission coefficient (b) of the antenna ports labelled in Fig. 5.

First, we compare the reflection and transmission coefficients of antenna elements labelled in Fig. 5. The most distorted elements are connected to the blue irregular vertex, and some of the more regular elements are connected to the red vertex. Fig. 6 plots the simulated reflection coefficient and coupling to nearest neighbors for the antenna elements listed in Fig. 5. Note that the reflection coefficient of a single port is higher than the active reflection coefficient, as is typical for UWB arrays. In general, there is close agreement between the input impedance and coupling, for all ports. This agreement suggests that distorting the antennas to conform to the doubly curved surface has a minimal impact on performance.

Next, we evaluate the radiation patterns. The array is excited to generate a right-handed circularly polarized beam. The weights feeding each port are calculated by illuminating the array with an incident right-handed circularly polarized plane wave and noting the received complex voltage at each element. Then, the array is excited with the complex conjugate of the received voltages and resulting the radiation patterns are calculated. The array can also radiate linear polarization, but circular is chosen here because it has a more intuitive definition when scanning over a very wide field of view [35]. In the future, more advanced beamforming approaches applicable to conformal arrays could be considered for increased pattern control [36].

Fig. 7 plots the simulated radiation patterns at 2, 5, and 10 GHz when the array points toward $\theta = 0°$. The radiation pattern is plotted on a modified coordinate system labelled $\theta_x$ and $\phi_x$, which has the $x$-axis pointing in the same direction as the main beam. This modified coordinate system provides a more intuitive visual representation of the beam because the main beam is circular when it is located at $\theta_x = 90°, \phi_x = 0°$. The irregular sidelobes at 10 GHz are expected since the array is under-sampled at frequencies above 4.75 GHz. We don't expect excellent cross-pol levels away from the scan direction because Vivaldi radiators have notoriously high cross-polarized radiation in the D-plane.

Fig. 8 plots the radiation pattern across the array's field of view at 2, 5, and 10 GHz. The beam is scanned between $\theta = -120°$ and $+120°$ every $30°$ in the $\phi = 0°$ plane. For reference, the dashed lines correspond to the theoretical gain of hemispherical $\left(\frac{1}{2} + \frac{\cos(\theta)}{2}\right)$ and planar arrays $(\cos(\theta))$. The array generates well-formed beams at the various frequencies and scan angles. Sidelobe and cross-pol levels are commensurate with planar arrays. The gain vs. scan angle generally follows the theoretical value of a hemispherical array. The gain at wide scan angles at 2 GHz is significantly larger than the theoretical value based on projected area. This is because theory assumes the array is electrically large ($r \gg \lambda$), but this assumption is not valid at low frequencies such as 2 GHz ($r = \lambda/1.7$).

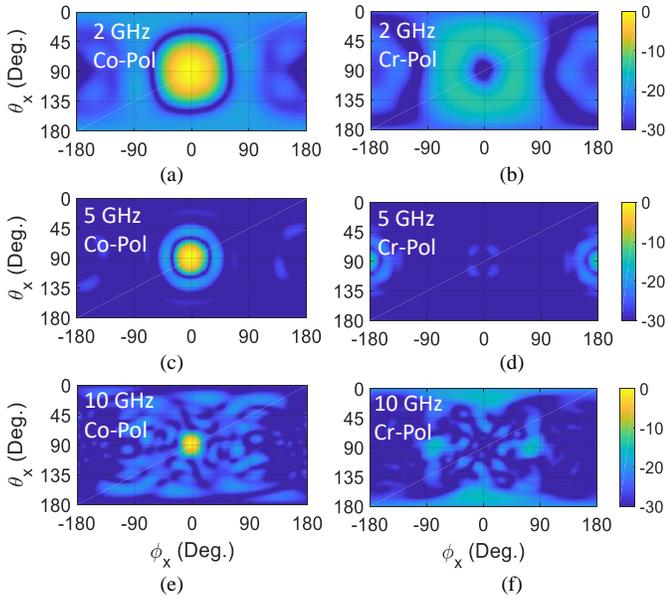

Fig. 7. Simulated radiation patterns when the array points toward $\theta = 0°$. Patterns are plotted on a modified coordinate system $\theta_x$ and $\phi_x$, which has the $x$-axis pointing in the same direction as the main beam.

Fig. 9 plots simulated the gain, loss, cross-pol, and peak reflection vs. frequency at different elevation scan angles. The realized gain is the product of the antenna gain and mismatch loss. For each elevation angle ($\theta$), the array is scanned over all azimuth ($\phi$) angles. The linewidth of the curves in Fig. 9 (a)-(c) correspond to $\pm 1$ standard deviation across azimuth. In general, the linewidth is less than 0.5 dB in Fig. 9(a) which suggests the gain is relatively independent of azimuth scan angle as expected. The dashed lines in (a) plot the gain of a theoretical hemispherical antenna with a 100% aperture efficiency and the same 181.5 mm diameter. Simulations show the array achieves close to optimal performance. As mentioned earlier, the theoretical gain is constant above 4.75 GHz because the array is undersampled at these frequencies. Again, theory assumes a large projected area ($A_p \gg \lambda^2$) which is less valid at lower frequencies and wider scan angles. Therefore, we observe more discrepancy between theory and simulation in such regimes.

The loss in Fig. 9(b) corresponds to the ratio of realized gain to directivity, which is identical to the product of the mismatch loss and radiation efficiency. The loss is dominated by the mismatch loss since each element achieves $> 95\%$ radiation efficiency. The mismatch loss is around 2 dB in the frequency range 1.5-5 GHz, which is significant but expected. The mismatch loss could likely be improved in future arrays that optimize the element impedance match. The mismatch loss is

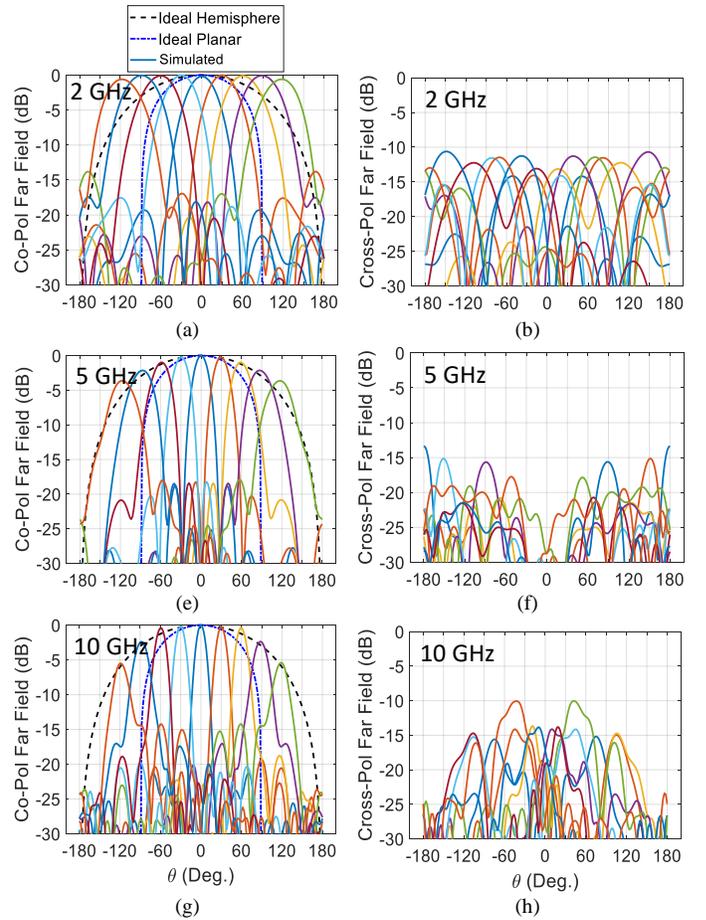

Fig. 8. Simulated gain when the array is scanned to different elevation angles in the $\phi = 0°$ plane. The patterns are normalized by the peak gain of the $\theta = 0°$ case.

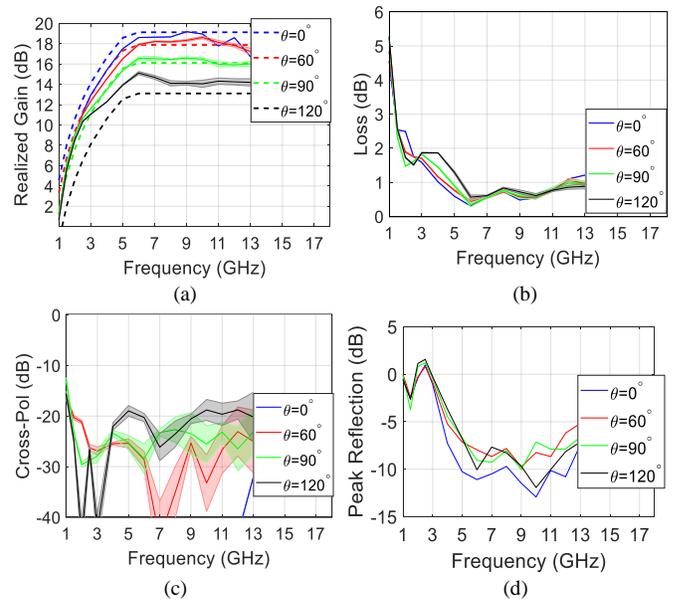

Fig. 9. Simulated performance. (a) Realized gain when the beam points toward different elevation angles. (b) Combination of mismatch and radiation efficiency loss. (c) Cross-pol in the scan direction. (d) Peak active reflection coefficient. In all cases, the array is scanned to every azimuth angle. In (a)-(c), the linewidth corresponds to $\pm 1$ standard deviation across azimuth.

< 1.3 dB above 5 GHz and scan angles less than 120°. The cross-pol in Fig. 9(c) corresponds to the ratio of left-handed circular polarization to right-handed circular polarization in the scan direction. In general, the cross-pol is relatively low (< -20 dB) even at wide scan angles.

Fig. 9(d) plots the worst-case active reflection coefficient for all azimuth scan angles and all antenna ports. For example, if every element is excited with 0 dBm or less power and the elevation angle is $\theta = 60°$, a peak reflection of -5 dB would mean that all elements have ≤-5 dBm power reflected into their ports when the array is scanned to any azimuth angle. At frequencies less than 3 GHz, a handful of elements have high peak reflection near 0 dB even though the overall mismatch loss is around 2 dB. A 0 dB reflection could be problematic if the array transmits high power and the power amplifiers cannot handle high reflection. Other methods of beamforming that account for the antenna array scattering parameters might help mitigate this high peak reflection in the future.

## V. Fabrication and Measurement

The fabricated array is shown in Fig. 10(a). The array is constructed by screwing 20 separate modules together. Two of the modules are shown in Fig. 10(b). Each module is 3D printed with titanium ($Ti_6Al_4V$) using the GE Additive Concept Laser M2, which can print parts up to 245mm x 245mm x 330mm in size. Many factors affect cost such as size, weight, and structural support removal time. The overall cost for printing the 20 modules from a commercial vendor is roughly $9k (USD) which translates into a price/element of $173 (i.e., $86/port).

Each element is fed with an SMP connector that is 3D printed with the antenna. These connectors are precisely fabricated so that commercial female SMP connectors can mechanically snap into the socket while also ensuring there is good electrical contact. A detent in the connector helps ensure a good connection is maintained if there is some vibration or stress on the input cables. Additional information regarding 3D printing RF push-on-connectors with the antenna can be found in [10].

Fig. 10(c) shows the experimental setup of the array on the antenna positioner. The array is mounted to a roll over azimuth far field antenna measurement system which allows for characterizing the entire 3D radiation pattern. The measurements are calibrated using the gain transfer method by measuring the gain of a known reference horn antenna. The measurement system is calibrated to the antenna connectors which removes the loss of the RF cables and switches. The array is characterized by measuring the complex embedded element pattern of all 104 antenna ports and then using digital beamforming to post process the antenna array patterns. Each low-gain antenna element is measured in azimuth from $\phi = 0°$ to 360° with 7.5° spacing and in elevation from $\theta = 0°$ to 180° every 7.5°. Time domain gating with a 500 mm (1.7 ns) wide window is employed to help reduce the impact of reflections from antenna positioner, feed cables, and chamber walls. Furthermore, we employed a spatial filtering routine that decomposes the far field into the spherical harmonics that are supported by the 185 mm diameter sphere [37]. This helps filter out unphysical far field oscillations that cannot be excited by the finite sized hemispherical antenna [38]. In addition, decomposing the far field into spherical harmonics allows for accurate interpolation of the far field on a grid with 2° spacing in azimuth and elevation. Measuring the 3D radiation patterns of all 104 ports within a timely manner is made possible by an absorptive single pole 36 throw switching matrix that measures 36 antenna ports at every angular position. Therefore, 3 scans are necessary to measure every antenna port. All antenna ports that are not connected to the switching matrix are terminated with 50 Ω loads.

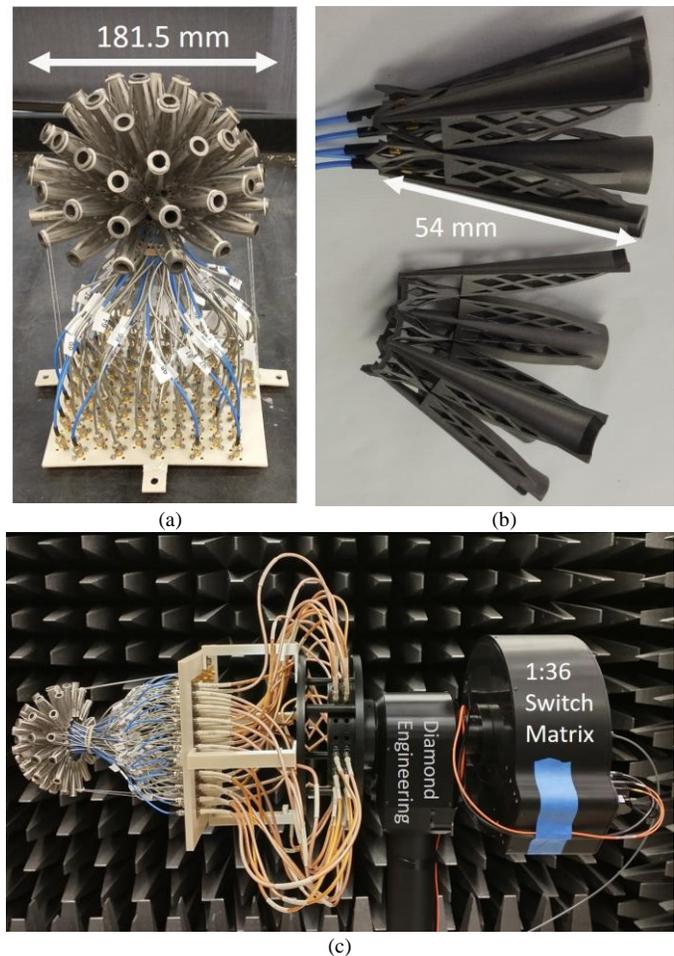

Fig. 10. (a) Fabricated array. (b) Two of the 20 modules comprising the array. (c) Array mounted on the spherical far field antenna measurement system.

As in simulation, beamforming at a given angle is accomplished by simply complex conjugating the received complex voltages at every port. This method of beamforming requires measuring and storing the complex far field at every angle. This corresponds to 104 ports × 101 frequencies × 49 azimuth angles × 25 elevation angles = $13 \times 10^6$ complex values, which could be a challenging amount of data to deal with for applications requiring real-time beamforming. In the future, more elegant beamforming techniques could be developed such as using an analytic model for the embedded element patterns. Furthermore, it is possible to accurately compress the stored data using a coupling matrix model [39].

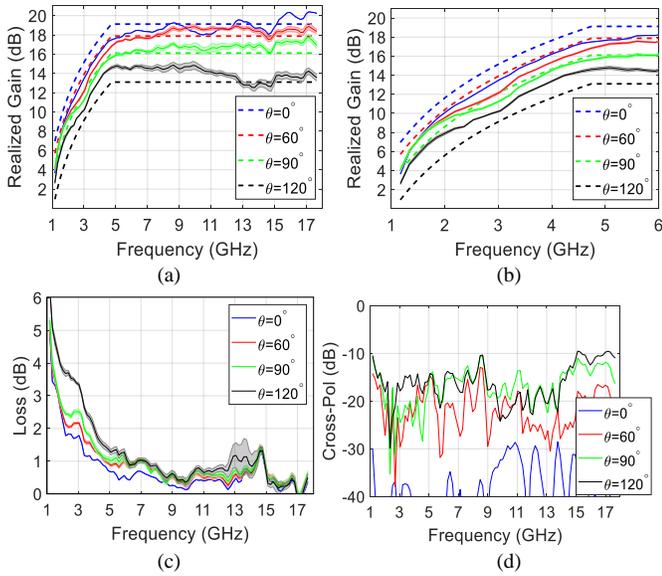

Fig. 11. Measured performance. (a) Realized gain when the beam points toward different elevation angles. (b) Zoomed in view of the gain from 1-6 GHz. (c) Combination of mismatch and radiation efficiency loss. (d) Cross-pol in the scan direction. In all cases, the array is scanned to every azimuth angle. In (a)-(c), the linewidth correspond to ±1 standard deviation across azimuth.

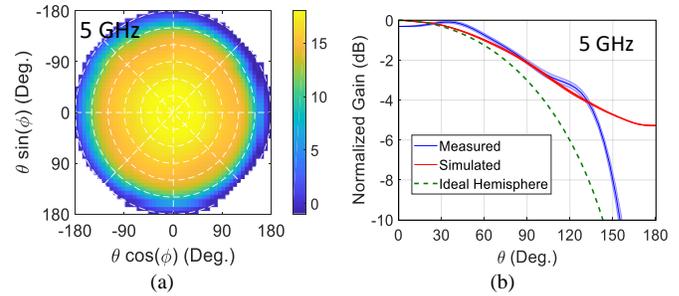

Fig. 12. Measured realized gain vs scan angle at 5 GHz. (a) Gain vs. scan angle in the $u$, $v$ coordinate system. (b) Average gain vs. elevation angle. The linewidth corresponds to the standard deviation across azimuth.

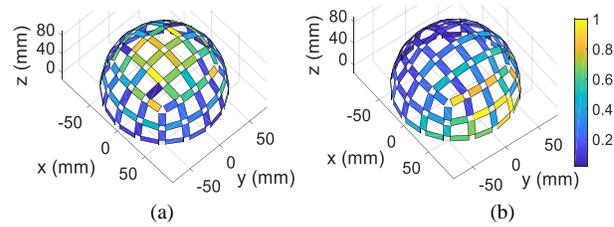

Fig. 13. Amplitude of the incident voltage that excites each element of the array at 5 GHz when the array points toward the $z$-axis (a) and $x$-axis (b).

Fig. 11(a) plots the average measured realized gain from 1-18 GHz and Fig. 11(b) zooms in on the grating lobe free band of 1-4.75 GHz. Fig. 11(c) plots the total loss which is the ratio of the gain to directivity. As in simulation, for each elevation angle, the beam is scanned to all azimuth angles ($\phi = -180° ... 180°$) and the gain is noted. Again, the linewidth at a given frequency in Fig. 11 (a)-(c) correspond to ±1 standard deviation in the gain/loss across all azimuth angles. The measured realized gain at broadside is generally within 2 dB of theory.

Fig. 11(d) plots the average cross-pol in the scan direction. The cross-pol at each point in the plot is averaged across all azimuth scan angles. The cross-pol is moderate with a value < -15 dB across much of the operating bandwidth and scan volume. This is higher than simulation though, which could be due to several sources. The imperfect electrical connections at the seams between the 20 modules that comprise the array could generate higher cross-polarization. In addition, scattering from the 104 coax cables feeding the antenna elements plus the 36 cables connected between the array and switching matrix can increase the cross-pol level.

To illustrate the large field of view of the array, Fig. 12(a) plots the gain at 5 GHz when the array points toward various scan angles. The $x$ and $y$ axes correspond to the $u$, $v$ coordinate system (i.e., $k_x d, k_y d$ coordinate system). There is a uniform gain vs. azimuth angle. The gain decreases at wide elevation angles in accordance with theory. Fig. 12(b) plots the average gain vs. elevation angle. Again, the linewidth corresponds to the standard deviation across azimuth. The measurement and simulation agree closely except when $\theta > 140°$. This discrepancy is due to the fact that the antenna positioner system and RF cables sit between the antenna under test and the reference antenna in this scan region.

Fig. 13 plots the amplitude of the incident voltage that excites each element of the array at 5 GHz when the array points toward the $z$-axis (a) and $x$-axis (b). Intuitively, the elements closest to the scan direction have the largest amplitude.

The co- and cross-polarized 3D radiation patterns at 2, 5, and 10 GHz are plotted in Fig. 14. All patterns correspond to the array pointing toward $\theta = 0°$. As in simulation, the patterns are plotted in a modified coordinate system $\theta_x$, $\phi_x$ with $x$-axis that points in the direction of the main beam. In general, there is decent agreement between the measured and simulated patterns.

Fig. 15 plots elevation and azimuth cuts of the radiation pattern at 2, 5, and 10 GHz. The left column of subfigures in Fig. 15 (i.e., (a), (c), and (e)) plot the case where the main beam points toward $\theta = 0°$, whereas the right column ((b), (d), and (f)) plots the case where the main beam points toward $\theta = 90°$, $\phi = 0°$. The sidelobe and cross-pol levels are quite moderate (< -15 dB) even when the array scans to $\theta = 90°$. The higher sidelobe levels at 10 GHz are due to pseudo-grating lobes since the element spacing is close to $1\lambda$ at this frequency. We refer to these sidelobes as 'pseudo-grating lobes' because the array is not truly periodic, which helps lower the sidelobes compared to the planar array case. As the frequency increases beyond 10 GHz, the sidelobes increase further due to the sparse array grid. For example, at 18 GHz the peak cross polarized sidelobe level is $-9$ dB when the beam points to $\theta = 0°$.

Fig. 16 plots the normalized co- and cross-polarized radiation patterns when the beam is scanned between $\theta = -120°$ and $+120°$ every $30°$ in the $\phi = 0°$ plane. There is decent agreement with simulations in Fig. 8. Dashed lines corresponding to the theoretical gain based on projected area of a hemisphere and planar array are also plotted. The peak gain at the different scan angles follows the theoretical value, which offers significantly wider scan volumes than a planar array. As in simulation, the measured gain at wider scan angles is larger than the theoretical value because the array is not very large.

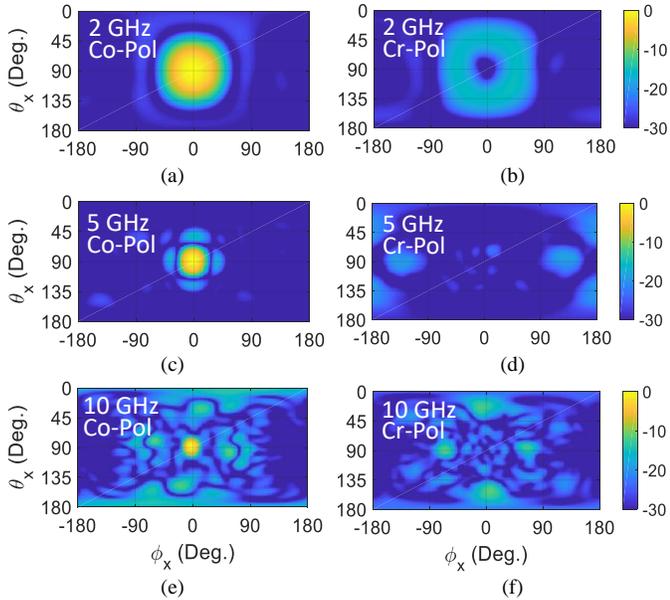

Fig. 14. Measured radiation patterns when the array points toward $\theta = 0°$. Patterns are plotted on a modified corrdinate system $\theta_x$ and $\phi_x$.

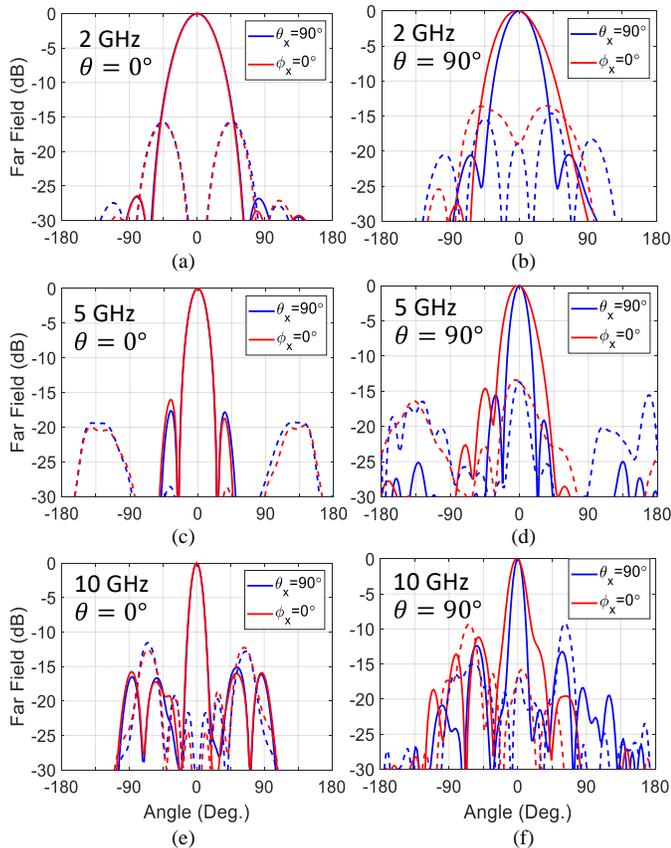

Fig. 15. Measured patterns in the $\theta_x = 0°$ and $\phi_x = 0°$ scan planes when the main beam points toward the $z$-axis (a), (c), (e), and $x$-axis (b), (d), (f). Solid lines are co-pol and dashed lines are cross pol.

Table I summarizes the simulated and measured array performance metrics. We define the operating frequencies to be when the total loss (product of mismatch loss and radiation efficiency) averaged over all azimuth angles is less than 2 dB. The maximum operating frequency is larger than we measured (>18 GHz) or simulated (>13 GHz) and could not be exactly determined. The loss and cross-polarization are averaged over all azimuth angles and frequencies on a linear scale within the operating bandwidth, and then converted to dB. It should be noted that the diameter of the simulated array is 9% smaller than the fabricated array. We did not rerun the simulations for the same array diameter as fabrication because simulations were only intended to provide a performance estimate, and these finite array simulations take a long time to run. The 1 dB difference between the measured and simulated peak gain is likely due to a combination of measurement error and the inaccuracy in the approximate array model for simulation

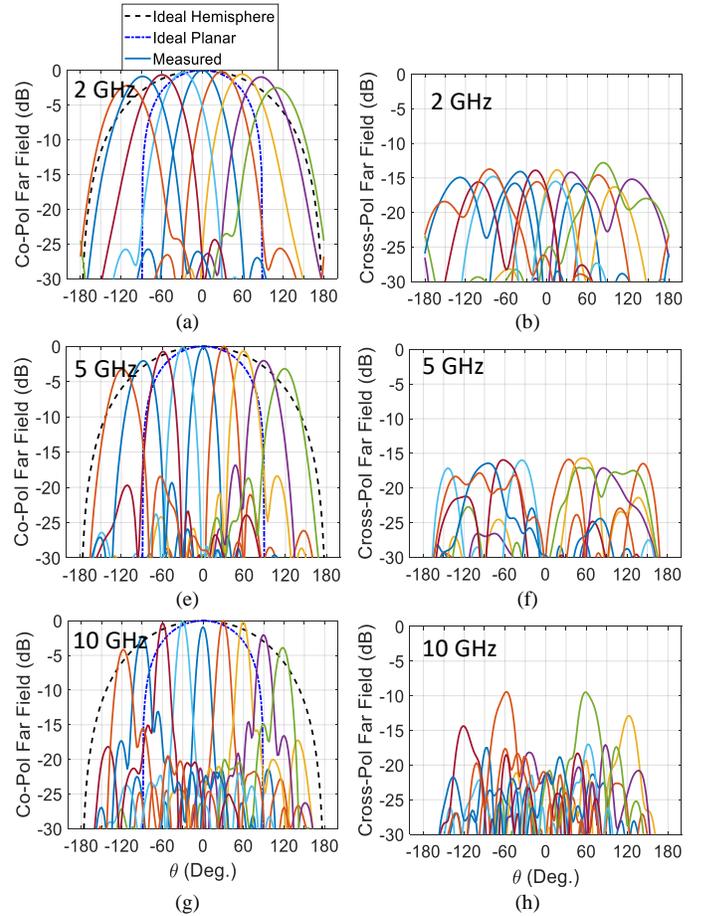

Fig. 16. Measured gain when the array is scanned to different elevation angles in the $\phi = 0°$ plane. The patterns are normalized by the peak gain of the $\theta = 0°$ case.

## VI. CONCLUSION

We report the first UWB antenna array on a doubly curved surface for wide angle scanning. We employ a quadrilateral meshing technique that generates a relatively uniform square lattice geometry. This geometry also supports the high coupling between antenna elements that is required for multi-octave bandwidths. The mapping approach is very general and can be applied to an arbitrary geometry. We then introduce the Vivaldi antenna element geometry that can be fabricated using a metal 3D printer. SMP connectors are integrated into the antenna elements, which significantly simplifies assembly. A proof-of-concept UWB array covering the surface of a hemisphere is then demonstrated. Simulations and measurements show the array can generate well-formed beams at scan angles out to 120° from the z-axis (i.e., $3\pi$ steradians) from 2 GHz to 18 GHz. The measured gain is within 2 dB of the simulated and theoretical values at all frequencies and scan angles.

TABLE I
SIMULATED AND MEASURED ARRAY PROPERTIES

|  | Simulation | Measurement |
|---|---|---|
| Diameter | 161.5 mm | 181.5 mm |
| Polarization | Dual-Linear | Dual-Linear |
| Grating Lobe Free | < 5.34 GHz | < 4.75 GHz |
| Freq. Range ($\theta = 0°$) | (2.3 GHz, >13 GHz) | (2.1 GHz, >18 GHz) |
| Freq. Range ($\theta = 90°$) | (1.7 GHz, >13 GHz) | (3.4 GHz, >18 GHz) |
| Avg. Loss ($\theta = 0°$) | 0.9 dB | 0.6 dB |
| Avg. Loss ($\theta = 90°$) | 0.9 dB | 0.8 dB |
| Avg. X-Pol ($\theta = 0°$) | -42 dB | -35 dB |
| Avg. X-Pol ($\theta = 90°$) | -24 dB | -15 dB |
| Peak Realized Gain ($\theta = 0°$) | 19.2 dB | 20.4 dB |
| Peak Realized Gain ($\theta = 90°$) | 16.9 dB | 18.3 dB |
| Peak Directivity ($\theta = 0°$) | 19.7 dB | 20.7 dB |
| Peak Directivity ($\theta = 90°$) | 17.5 dB | 18.3 dB |

This work is intended to serve as a baseline estimate for the performance of future UWB, wide scan arrays employing tightly coupled antenna elements. The current hemispherical prototype is only 52 elements in size. Larger arrays will generally have larger radii of curvature and more uniform lattices that make optimizing their performance more straightforward. Another issue with the current prototype is there is an imperfect electrical contact between the 20 modules that comprise the array. We expect these seams between modules to degrade cross-pol and impedance match, but it is unclear at this moment how significant this performance degradation is. A natural extension of this work is to consider more advanced UWB radiating elements such as a tightly coupled dipole array. The dipole array could achieve a similar impedance bandwidth as Vivaldi elements while reducing cross-polarized radiation. In addition, the dipole array has a significantly lower profile than a Vivaldi array, which would allow for realizing a smaller radius of curvature. In this work, we used a relatively crude beamforming approach based on complex conjugation. In the future, more elaborate pattern synthesis techniques should be considered to control parameters such as cross-polarized radiation, sidelobe level, and null placement. Developing accurate analytic models for the embedded element patterns would also aid beamforming. This further motivates development of low-profile conformal antenna elements because they have a simpler and more accurate analytic model than electrically large Vivaldi elements.

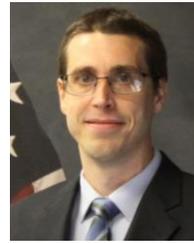

**Carl Pfeiffer** received the B.S.E., M.S.E. and Ph.D. degrees in electrical engineering from The University of Michigan at Ann Arbor, Ann Arbor, MI, USA, in 2009, 2011, and 2015 respectively.

He was an electrical engineer at Defense Engineering Corp. from 2016 to 2021 and is currently a senior research electronics engineer with the Air Force Research Laboratory at Wright-Patterson Air Force Base, OH, USA. His research interests include phase arrays, engineered electromagnetic structures (metamaterials, metasurfaces, frequency selective surfaces), antennas, microwave circuits, plasmonics, optics, and analytical electromagnetics/optics.

Dr. Pfeiffer has served as a technical program committee member of EUCAP 2017 and has received top reviewer awards for IEEE Trans. Antennas Propag. in 2019 and 2020. He received the Kittyhawk AOC Research and Technology Development Award in 2020.

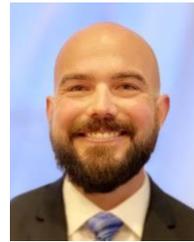

**Jeff Massman** received the B.S.E. and M.S.E. degrees in electrical engineering from United States Air Force Academy and Air Force Institute of Technology in 2008 and 2010, respectively.

He is currently the lead for the antenna and electromagnetic structures lab team with the Multiband Multifunction RF Sensing Branch of the Sensors Directorate Air Force Research Laboratory at Wright-Patterson Air Force Base, OH, USA and a PhD student at the Air Force Institute of Technology. His research interests include additive manufactured antennas, phased arrays, conformal frequency selective surfaces, and simultaneous transmit and receive devices.